\documentclass[useAMS,usenatbib]{mn2e}

\usepackage{graphicx}
\usepackage{amsmath}
\usepackage{amssymb}

\title[Gamma-rays from the IC $e^\pm$ pair cascade in Cen~A]
{Gamma-rays from the IC $e^\pm$ pair cascade in the radiation field of an accretion disk:
Application to Cen~A}

\author[J. Sitarek and W. Bednarek]{J. Sitarek$^{1,2}$
and W. Bednarek$^{2}$ \\
$^{1}$Max-Planck-Institut f\"ur Physik, D-80805 M\"unchen, Germany, jsitarek@mppmu.mpg.de\\
$^{2}$University of \L\'od\'z, PL-90236 \L \'od\'z, Poland, bednar@astro.phys.uni.lodz.pl}

\begin{document}

\date{Accepted . Received ; in original form }

\pagerange{\pageref{firstpage}--\pageref{lastpage}} \pubyear{2007}

\maketitle

\label{firstpage}

\begin{abstract}
The very short time scale variability of TeV $\gamma$-ray emission from active galaxies suggests that the acceleration process of particles and the production of primary $\gamma$-rays likely occurs relatively close to the accretion disk. We calculate the $\gamma$-ray spectra produced in an Inverse Compton $e^\pm$ pair cascade initiated by primary $\gamma$-rays which are injected close to the surface of the accretion disk. Possible synchrotron energy losses of secondary cascade $e^\pm$ pairs are also taken into account.
Since the soft radiation field is anisotropic, the resulting $\gamma$-ray spectra strongly depend on the observation angle. We investigate their basic properties for different parameters describing such a model. The model is applied to the misaligned blazar Cen~A recently detected in the TeV $\gamma$-rays. We conclude on the site of the $\gamma$-ray emission region in Cen~A based on the comparison of the model with the observations of this source in the GeV-TeV energy range.  
\end{abstract}

\section{Introduction}

Up to now, the GeV-TeV $\gamma$-ray emission has been observed from a few tens of active galactic nuclei.
It is widely believed that this radiation is emitted by particles accelerated in the  
jet launched from the inner part of an accretion disk around a super massive black hole.
Short time scale variability of this $\gamma$-ray emission (see e.g. recent results in the case of Mrk 501 
and PKS 2155-304,~Albert et al.~2007 and Aharonian et al.~2007) strongly suggest that the radiation originates in the inner jet close to the accretion disk. 
In such a case the radiation field is strong enough that $\gamma$-rays are absorbed and produced in the cascade process. In fact, importance of different cascade scenarios, as a possible sources of $\gamma$-rays from relativistic plasma, has been intensively explored during almost 30 years both numerically 
(e.g. Pozdnyakov, Sobol \& Sunyaev~1983, Coppi 1992) and analytically (e.g. Svensson, 1984, 1987). In the 90-ties, it became clear that $\gamma$-rays observed from active galaxies likely originate inside the jet. Therefore, the cascade processes were considered isotropically in the relativistic blob frame (e.g. Mannheim \& Biermann~1992,
Blandford \& Levinson~1995, Marcowith, Henri \& Pelletier~1995) or were assumed to develop mono-directionally in the general direction of the jet, i.e. perpendicular to the accretion disk (e.g. Bednarek \& Kirk~1995, Bednarek~1997, Bednarek \& Protheroe~1999). Note, that the $e^\pm$ pair cascade processes have been also extensively studied with the application to $\gamma$-ray burst sources (although at lower energies), where the large optical depths for $\gamma$-rays are expected.
In the present paper, in contrast to the previous works, we consider a more complicated scenario in which the production of $\gamma$-rays in the cascade process is treated three-dimensionally.
The primary \mbox{$\gamma$-rays}, produced close to the disk surface, can be severely absorbed in the disk radiation developing Inverse Compton (IC) $e^\pm$ pair cascade in the whole volume above the accretion disk. 
In this case the cascade $\gamma$-ray spectra depend on the observation angle in respect to the orientation of the accretion disk due to the anisotropic radiation from the disk surface.
Therefore, the role of the anisotropic radiation field of the accretion disk in the process of $\gamma$-ray production and propagation can be essential (see e.g. the recent calculations of the effects of $\gamma$-ray absorption in the case of two famous sources 3C 273 and 3C 279 in Sitarek \& Bednarek~2008). 

In this paper we analyze in detail the cascade initiated by the primary TeV $\gamma$-rays injected from a compact region (a blob) moving along the jet in the radiation field of the accretion disk. 
Such a cascade can develop in the whole volume above the disk, where the radiation field is strongly anisotropic. 
We calculate the cascade $\gamma$-ray spectra emerging at different directions in respect to the disk axis.
The possible role of the magnetic field above the accretion disk is also investigated.
As an example, we consider the nearby active galaxy, Cen~A which shows a jet aligned at a relatively large angle to the observer (see for review Israel~1998). This galaxy has been detected by the EGRET telescope in GeV energies with 
the spectrum well described by the single power law with the spectral index $\sim -2.4$~(Sreekumar et al. 1999). 
Recently, the {\it Fermi} Observatory also reported discovery of a weaker $\gamma$-ray source towards Cen~A, but with a much steeper (spectral index $\sim -2.9$) spectrum (Abdo et al.~2009).
Cen~A has been also detected by the H.E.S.S. telescopes above $\sim 250$ GeV~(Aharonian et al.~2009), becoming the second radio galaxy detected in VHE $\gamma$-rays. 
The source is weak, having a spectrum with the power law index $\sim -2.7$. 
This TeV spectrum fits nicely with the extrapolation of EGRET spectrum to the higher energies.
Therefore, it looks that the EGRET and H.E.S.S. observes Cen~A in a relatively high state in contrast
to the recent observations by the ${\it Fermi}$ Observatory. 

Note that some previous models, proposed as possible explanation of GeV-TeV $\gamma$-ray emission 
from Cen~A type misaligned blazars (e.g. Lenain et al.~2008, Ghisellini et al.~2005),
have not considered formation of this emission in the cascade processes developing in the whole volume above the accretion disk near the central engine of active galactic nuclei. 
Other models postulate production of TeV $\gamma$-rays in a pulsar-like cascade mechanism in the magnetosphere of the black hole (Neronov \& Aharonian~2007, Rieger \& Aharonian~2008).
Considered here scenario in which $\gamma$-rays initiate IC $e^\pm$ pair cascade in the anisotropic radiation field of the accretion disk naturally explains significant emission at relatively large angles to the direction of the jet expected to be aligned with the axis of the accretion disk around a super-massive black hole.

\section{Scenario for $\gamma$-ray production}

As we noted above, in contrast to previous calculations we indent to consider the production of $\gamma$-rays in the IC $e^\pm$ pair cascade developing in
the whole volume above the accretion disk. In such a case we have to treat the problem three-dimensionally due to the highly anisotropic radiation field seen by primary and secondary cascade particles. We assume that the primary $\gamma$-rays are produced in one of the popular models, e.g. synchrotron self-Compton (SSC) (e.g. Maraschi, Ghisellini \& Celloti~1992) or hadronic cascade model (Mannheim \& Biermann~1992). 
The volume of the injection region (the blob) is much smaller than the volume in which the cascade develops. Therefore, we assume the blob to be a point-like source of primary $\gamma$-rays. 
The blob moves mildly relativistically along the jet. In the case of Cen~A the velocity of the blob is estimated on $\beta\approx 0.5c$, where $c$ is the velocity of the light (Tingay et al.~1998, Hardcastle et al.~2003). The blob injects isotropically primary $\gamma$-rays (in the blob frame) at a specific distance $H$ from its base or throughout the range of distances starting from the base. However, the distribution of primary $\gamma$-rays in the disk frame is not isotropic due to the Doppler beaming effect defined by the velocity of the blob.

These primary $\gamma$-rays leave the blob/jet region and initiate the IC $e^\pm$ pair cascade in the anisotropic radiation field of the accretion disk. The cascade develops in the whole volume above the accretion disk. 
We assume that primary $\gamma$-rays are injected from the point like blob with the power law spectrum.  

Moreover, a magnetic field with a dipole structure exists above the accretion disk. 
The magnetic field may significantly re-distribute directions of cascade $e^\pm$ pairs which additionally suffer synchrotron energy losses. 
We apply following method for tracking $e^\pm$ pairs in the assumed dipole magnetic field above the accretion disk.
The movement of the electron/positron is a sum of two movements: along the magnetic field lines and circular around the field lines forming a spiral along the magnetic line.
If gyroradius is comparable or larger than other characteristic distances (like mean free path for IC, distance from the accretion disk, etc.), in each steep we 
calculate the new position of the particle after a fraction of the turn of the spiral. 
This approach is numerically very inefficient if gyroradius becomes much smaller than other characteristic distances. 
In this case the cascade code switches into tracking of the guiding center (the center of the spiral), which is moving along 
the magnetic field line. Since in this situation direction of the particle is 
constantly changing and the IC mean free path depends on the particle direction 
(with respect to the source of the radiation - accretion disk), we use values of 
mean free paths which are properly averaged over true directions of the tracked 
particle.

As a result of the Inverse Compton cooling in the radiation field of the accretion disk the energies of secondary $e^\pm$ pairs are reduced by the amount of energy taken by produced secondary $\gamma$-ray photon. 
Moreover, we include also the synchrotron energy losses of secondary $e^\pm$ pairs when considering the fate of secondary cascade $e^\pm$ pairs. All the above mentioned processes (absorption of $\gamma$-rays, inverse Compton scattering of disk radiation, magnetic deflection of $e^\pm$ pairs and their synchrotron energy losses) are taken into account in our Monte Carlo code which follows the development of such complicated anisotropic IC $e^\pm$ pair cascade. These radiation processes ($\gamma-\gamma\rightarrow e^\pm$ and Inverse Compton) are considered by applying full cross sections. Using this code, we calculate the $\gamma$-ray spectra escaping at an arbitrary angle in respect to the jet axis. 
Its application seems to be specially interesting in the case of the blazar which jet is aligned at a relatively large angle to the direction towards the observer (e.g. Cen~A or M87).

Note that defined above relatively simple geometrical scenario do not include many complications which might appear in the real situation such as possible interaction of the IC cascade $\gamma$-rays and $e^\pm$ pairs with the soft radiation produced in the source of primary $\gamma$-rays, in the synchrotron process by secondary $e^\pm$ pairs, or re-scattered by the matter surrounding the accretion disk. 
We also do not take into account possible annihilation of secondary, relativistic $e^\pm$ pairs. It can be easily estimated that for the observed $\gamma$-ray luminosities of misaligned blazars (the main objects of interest here) the density of
secondary $e^\pm$ pairs is too low for their efficient annihilation before significant energy losses on the IC and synchrotron processes.  
The complete three dimensional cascade model, which take into account all set of possible soft radiation fields and radiation processes, is too complicated at the present stage for numerical analysis. Additionally, it will require detailed and realistic knowledge on the content of the central engines of active galaxies which is at present not available. Therefore, modeling of simplified scenario in three dimensions considered in this paper gives already important progress in respect to 
previous works.

\section{Escape of $\gamma$-rays from the accretion disk radiation field}

We adopt a simple  optically thick and geometrically thin accretion disk model around super massive black hole as a dominant source of 
radiation in the central region of the active galactic nuclei (see Shakura \& Sunyaev~1973).
The emission of the accretion disk is treated as a black body with a power law temperature dependence on the distance, $r$, from the black hole, $T=T_{\rm in}(r/r_{in})^{-3/4}$, where $T_{\rm in}$ is the temperature at the inner disk radius $r_{\rm in}$. 

As an example, we apply the parameters expected for the misaligned blazar Cen~A which has been recently detected in TeV \mbox{$\gamma$-rays}~(Aharonian et al.~2009). 
It is supposed that the central engine of Cen~A harbors a super massive black hole with its mass estimated on $M_{\rm Cen A} = (5.5\pm 3.0)\times 10^7 M_\odot$~(Cappellari et al. 2009). 
A clear jet is observed in this radio galaxy. It is propagating at the angle towards the observer estimated on $\alpha_{jet}\approx 15^\circ-80^\circ$~(Horiuchi et al. 2006).
We apply the inner disk temperature equal to $T_{\rm in} = 3\times 10^4$~K, which is of the order of that one observed directly in other AGNs (e.g. 3C 273). 
The total disk luminosity with these parameters is estimated on 
$L_{\rm D} = 4\pi \sigma_{\rm SB}r_{\rm in}^2T_{\rm in}^4\approx 3.6\times 10^{41}$ erg s$^{-1}$, 
where $r_{\rm in} = 4.5\times 10^5M_{\rm Cen A}/M_\odot$ cm = $2.5\times 10^{13}$cm. 
Note that the disk is observed at a relatively large angle, which effectively reduces observed luminosity.
This accretion disk emission is additionally obscured by the circumnuclear disk. 
Therefore it can not be directly observed.

In the most general situation of the $\gamma$-ray photon injected at an arbitrary place above the accretion disk and at an arbitrary direction, the calculations are not straightforward since considered radiation field is highly anisotropic~(Carraminana~1992, Bednarek~1993). 
We investigate the optical depths for $\gamma$-rays as a function of their energies, injection place and injection angle for the radiation field defined by the parameters characteristic for Cen~A (see above).
Based on the above calculations, we determine the three-dimensional surfaces around the central engine of AGN at which the optical depths for $\gamma$-rays with specific energies, $E_\gamma$, and injection angles, $\alpha$ (measured in respect to the direction perpendicular to the disk surface), are equal to specific value. 
If this value is fixed on unity, then such surface is called the $\gamma$-sphere.
The shape of the $\gamma$-sphere can be in general quite complicated (see recent calculations for two
OVV blazars, 3C 273 and 3C 279, in  Sitarek \& Bednarek~2008.
However, evaluation of such a surface is very practical since $\gamma$-rays produced inside the $\gamma$-sphere are strongly absorbed, while those ones produced outside the $\gamma$-sphere can escape with a negligible absorption.

\begin{figure}
\centering
\includegraphics[scale=0.21, trim= 0 35 0 0, clip]{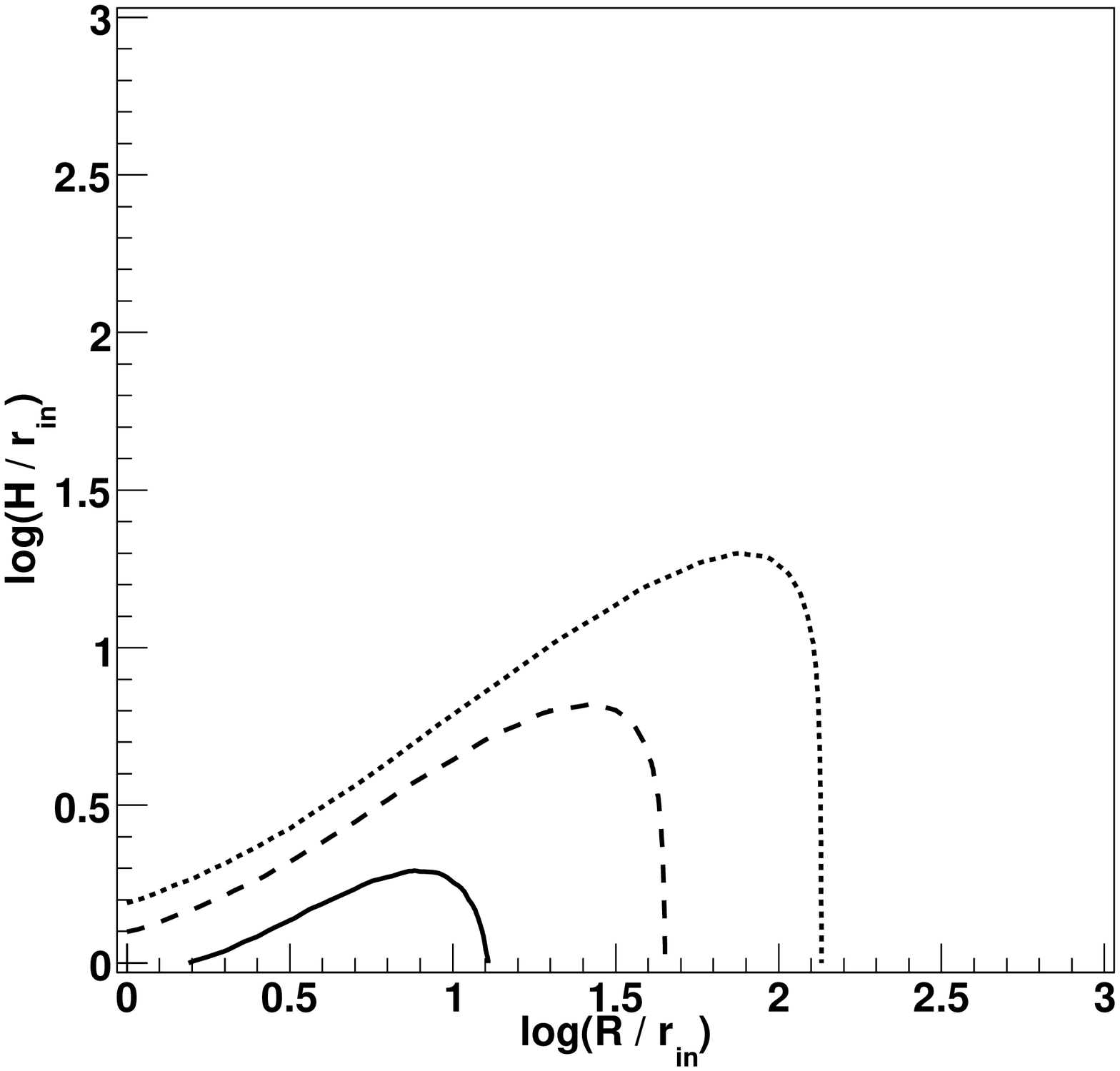} 
\includegraphics[scale=0.21, trim= 30 35 0 0, clip]{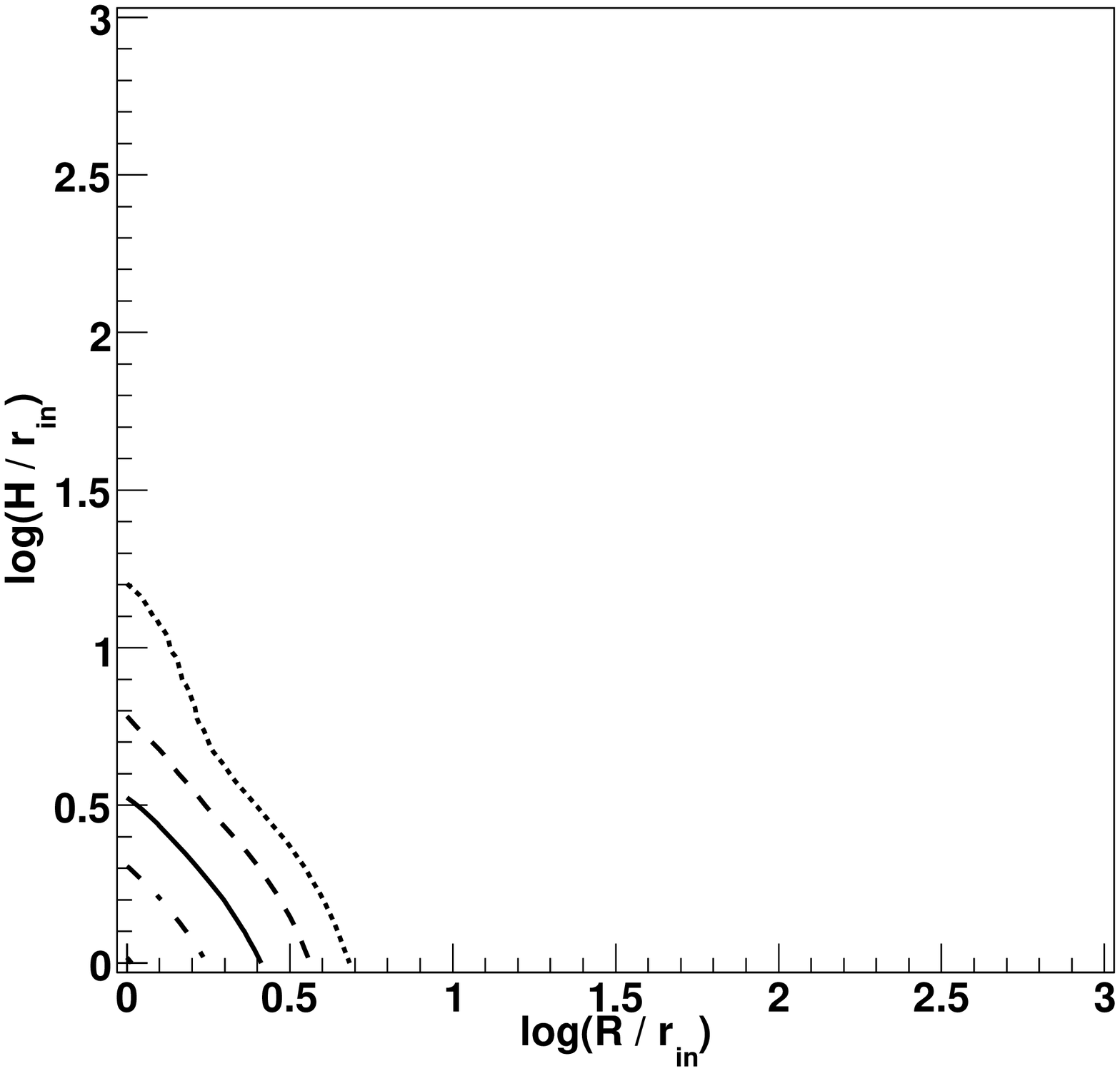} 

\includegraphics[scale=0.21, trim= 0 35 0 0, clip]{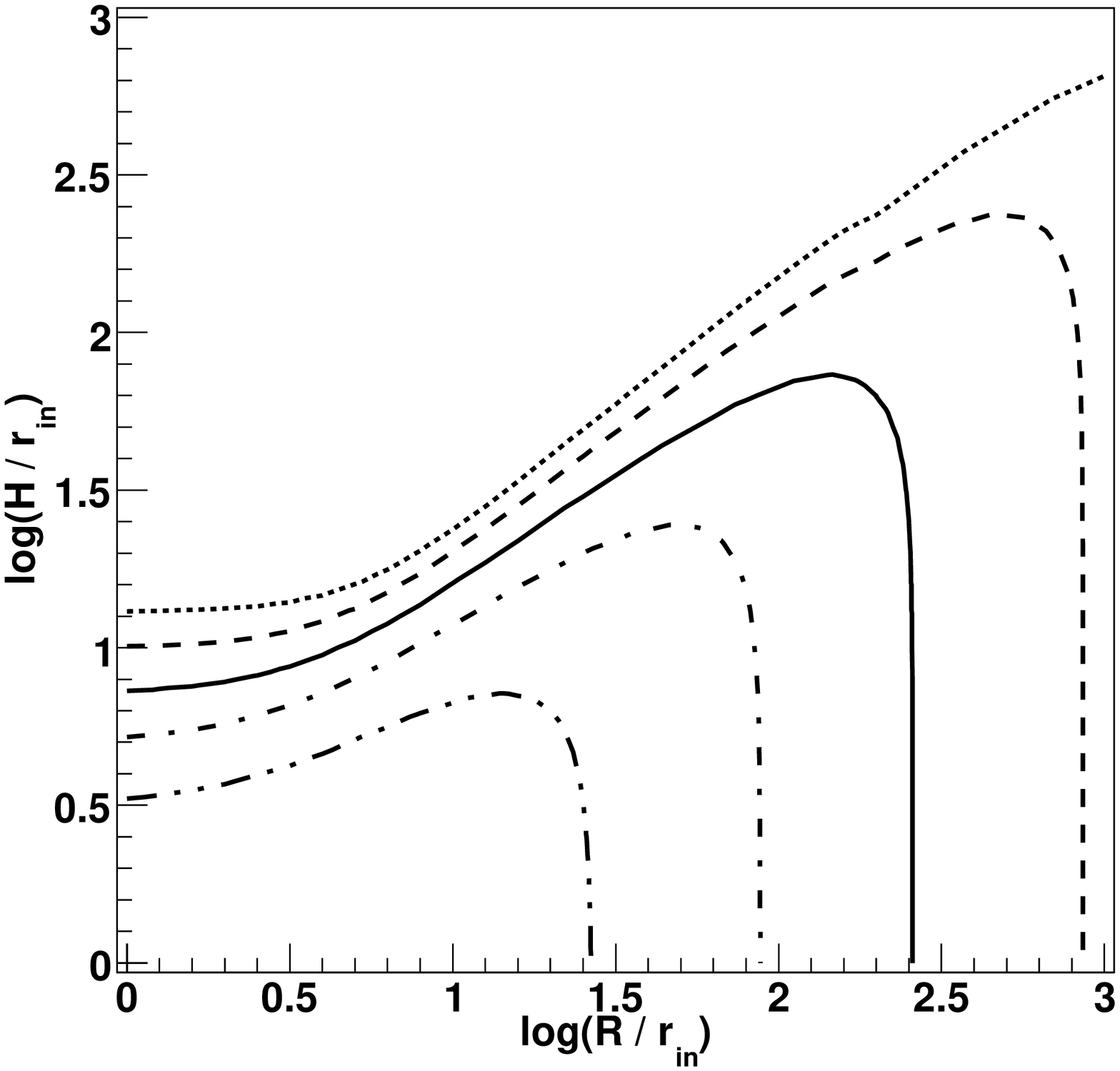} 
\includegraphics[scale=0.21, trim= 30 35 0 0, clip]{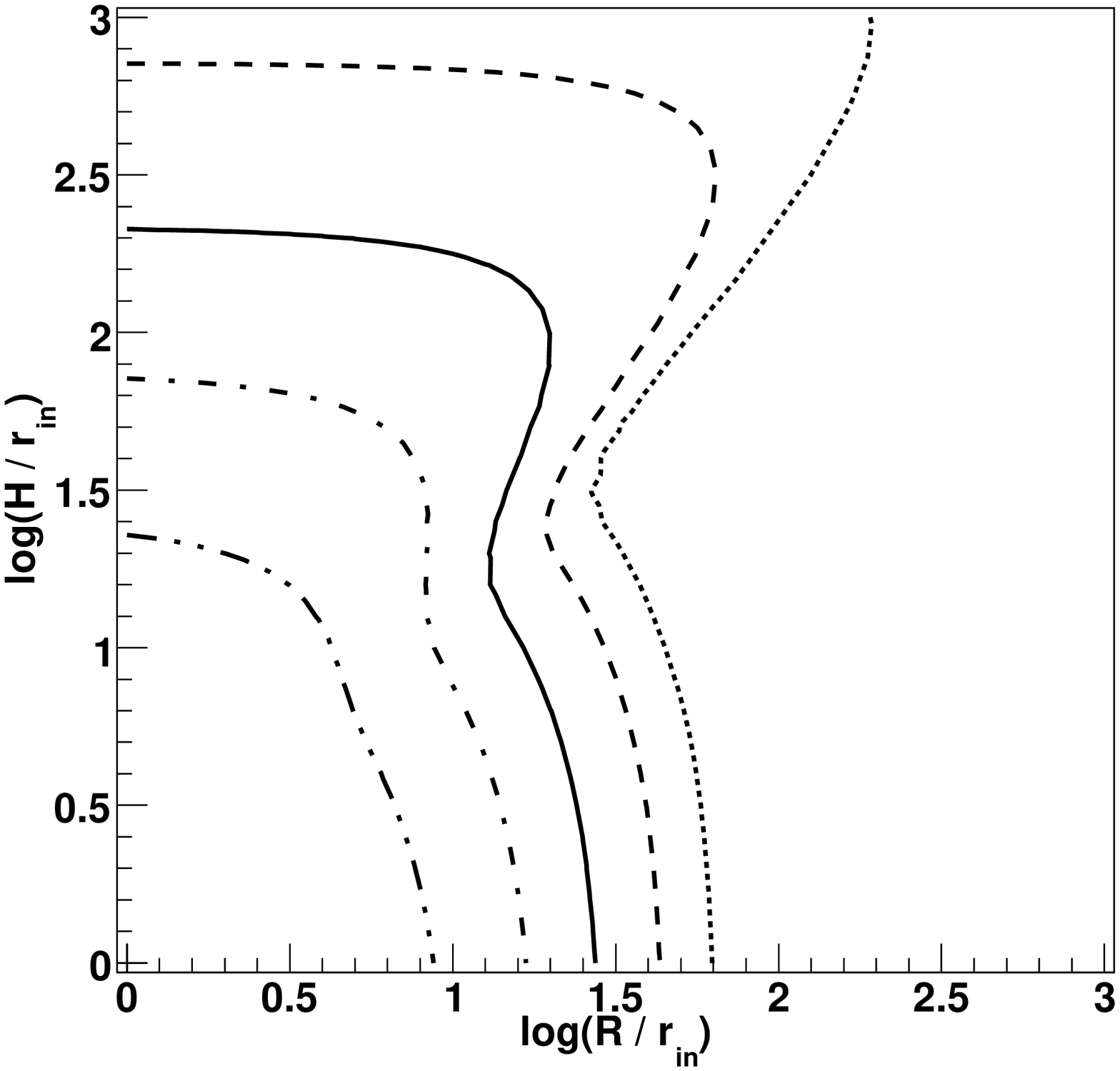} 

\includegraphics[scale=0.21, trim= 0 0 0 0, clip]{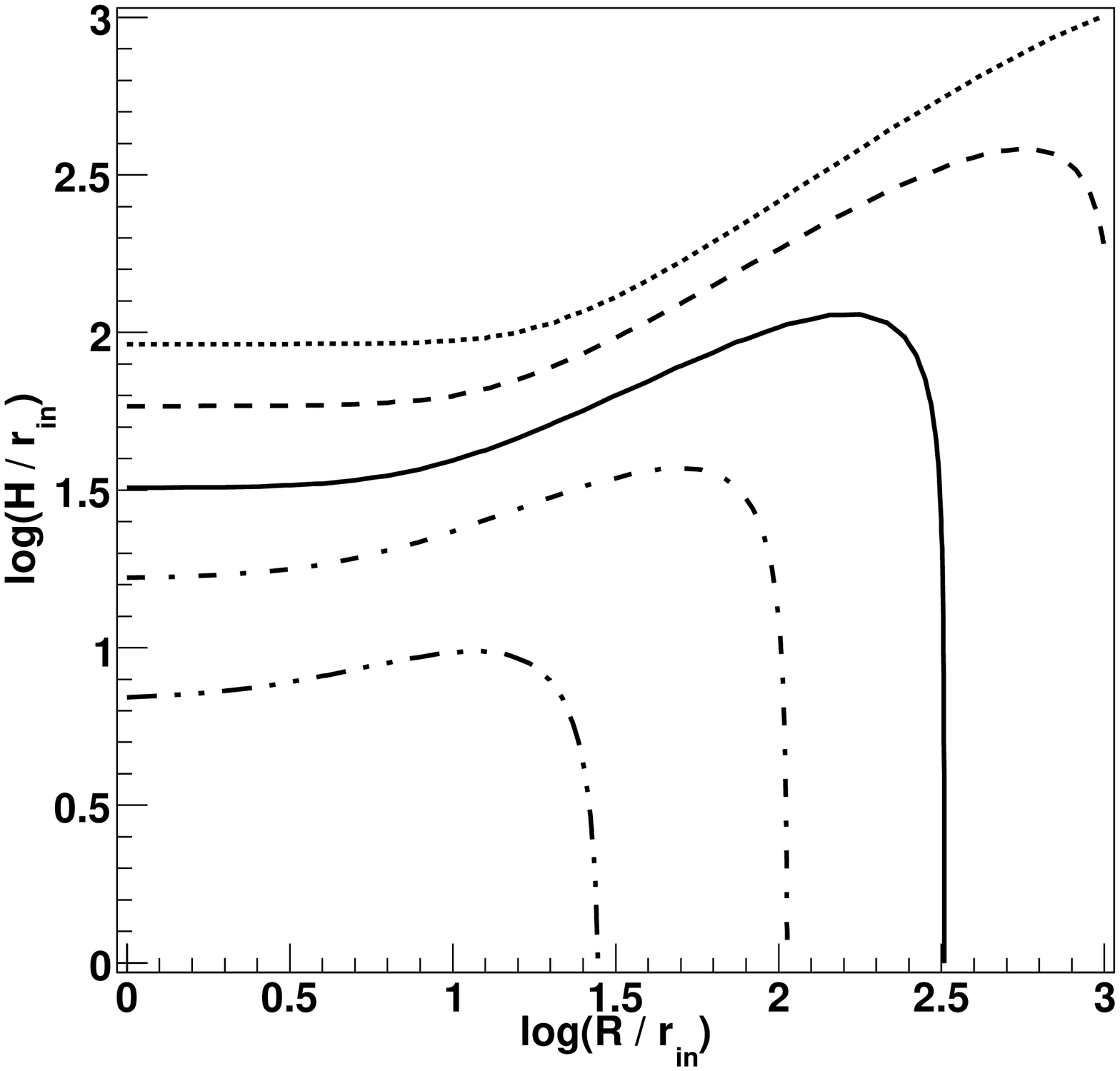} 
\includegraphics[scale=0.21, trim= 30 0 0 0, clip]{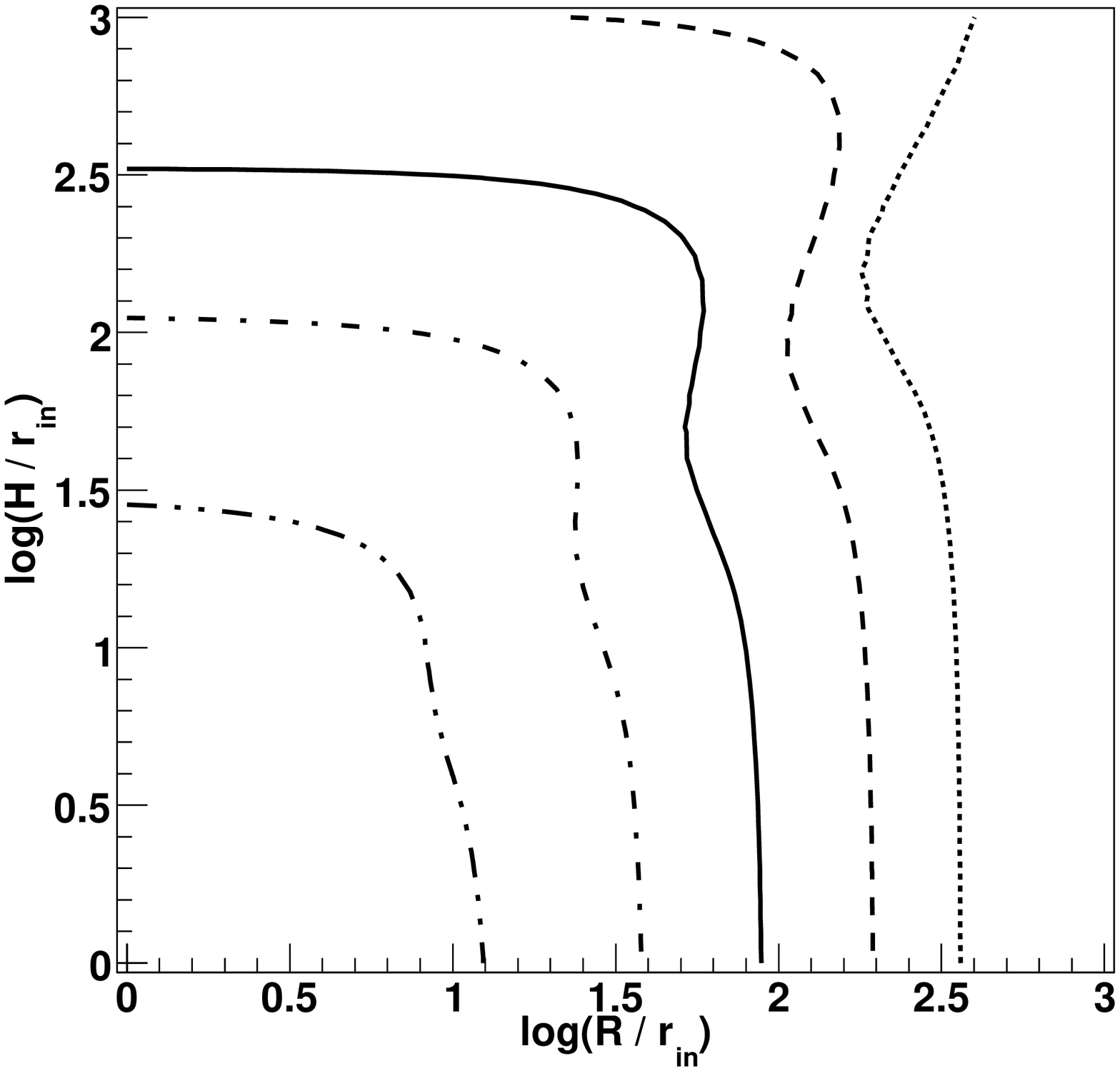} 
\caption{Three-dimensional $\gamma$-spheres around the accretion disk in Cen~A with the parameters mentioned in the text (H and R denotes the distances measured along and perpendicular to the jet, respectively). The $\gamma$-rays are injected at the angle to the jet axis 
equal to $\alpha = 0^\circ$ (left panels) and $60^\circ$ (right panels). 
The energies of $\gamma$-rays are equal to $E_\gamma = 0.1$ TeV (top figures), 1 TeV (middle), and 10 TeV (bottom).
Different curve styles represent lines of constant optical depth: $\tau=$ 0.1 (dotted), 0.3 (dashed), 1 (solid), 3 (dot-dashed) and 10 (dot-dot-dashed). The accretion disk in Cen~A is defined by $r_{in}=2.5\times 10^{13}$cm, $T_{in}=3\times 10^4$K.
}
\label{fig_gammasfery}
\end{figure}

The $\gamma$-spheres calculated for the parameters of the supposed accretion disk around the black hole in Cen~A are shown in Fig.~\ref{fig_gammasfery}. 
Specific figures show the $\gamma$-spheres for two values of the injection angle of the $\gamma$-rays 
$\alpha = 0^\circ$ and $60^\circ$ and three selected values of $\gamma$-ray energies: $E_\gamma = 0.1$, 1, and 10 TeV. 
Note that we are interested also in the optical depths for $\gamma$-rays propagating at large angle in respect to the disk axis. As we show latter, the 
cascade $\gamma$-rays can escape at such large angles in the case of jets moving with relatively low speeds (e.g. 0.5c as estimated in Cen~A).
The $\gamma$-spheres shows very interesting features.
For the injection angle of the $\gamma$-rays equal to $\alpha = 0^\circ$, the lowest values of the optical depths are along the axis of the accretion disk.
It is usually assumed that in this direction the jet is propagating. Thus, $\gamma$-rays 
injected along the jet axis have the highest probability of escape. Such is the case of the TeV
$\gamma$-ray sources observed from BL Lacs in which case the observation angle is typically
low. On the other hand, $\gamma$-rays injected at the same distance above the disk but farther from the jet axis suffer strong absorption. 
These general features are easy to understand if we keep in mind that most of the disk radiation is produced in the inner part of the accretion disk where the temperature is the largest. 
If $\gamma$-ray photon is moving along the jet, the $e^\pm$ pair production process in collisions with the disk radiation is strongly suppressed by an increasing energy threshold and a geometrical factor $1-\cos\theta$, where $\theta$ is the angle between the $\gamma$-ray and the low energy photon.
On the other hand if $\gamma$-rays are injected far from the jet (\mbox{$R \gtrsim H$}) the hot center of the accretion disk is seen by them at a larger angle $\theta$, which results in a stronger absorption. 
Note, moreover, the TeV $\gamma$-rays injected close to the disk surface meet very large optical depths, preventing their direct escape from the disk radiation. 
 
The optical depths for $\gamma$-rays injected at a large angle to the disk axis have different features
(see bottom panel in Fig.~\ref{fig_gammasfery}). 
TeV $\gamma$-rays produced in a jet even as far as $\sim 200 r_{in}$ can be efficiently absorbed.
The $\gamma$-spheres in this case have more complicated shape due to a weaker absorption of $\gamma$-rays for the directions fulfilling the condition $R/H \sim \tan 60^\circ $. 

Note that the results on the optical depths in the disk radiation can be easily generalized to the accretion disks with other inner temperatures and inner radii by
applying the general prescription described for the case of the optical depths inside massive binary systems (see section 4.2 in  Bednarek~2009). The calculated optical depths should be simply scaled by $R_{\rm in}$ and $T_{\rm in}^3$ and shifted in energy of $\gamma$-ray photons proportionally to $T_{\rm in}$.

\section{Cascade $\gamma$-ray spectra}

When calculating the $\gamma$-ray spectra escaping at a specific angle to the direction of the jet
we assume that primary $\gamma$-rays are injected isotropically from the moving region within the jet (the blob assumed point like in respect to the volume of the cascade) at the distance $H$ from its base.
 
\begin{figure*}
\centering
\includegraphics[scale=0.28, trim= 0  26 0 0, clip]{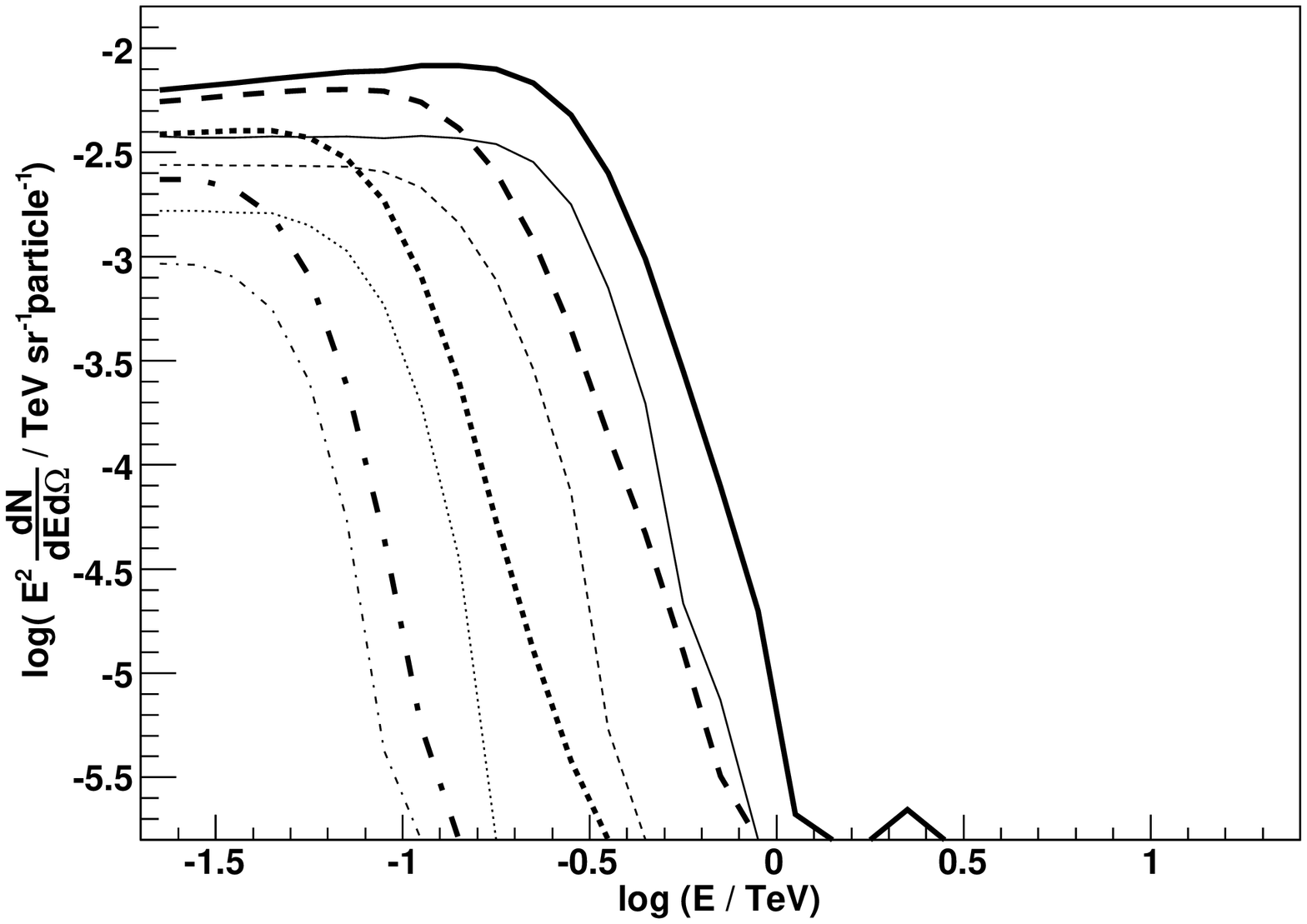} 
\includegraphics[scale=0.28, trim= 32 26 0 0, clip]{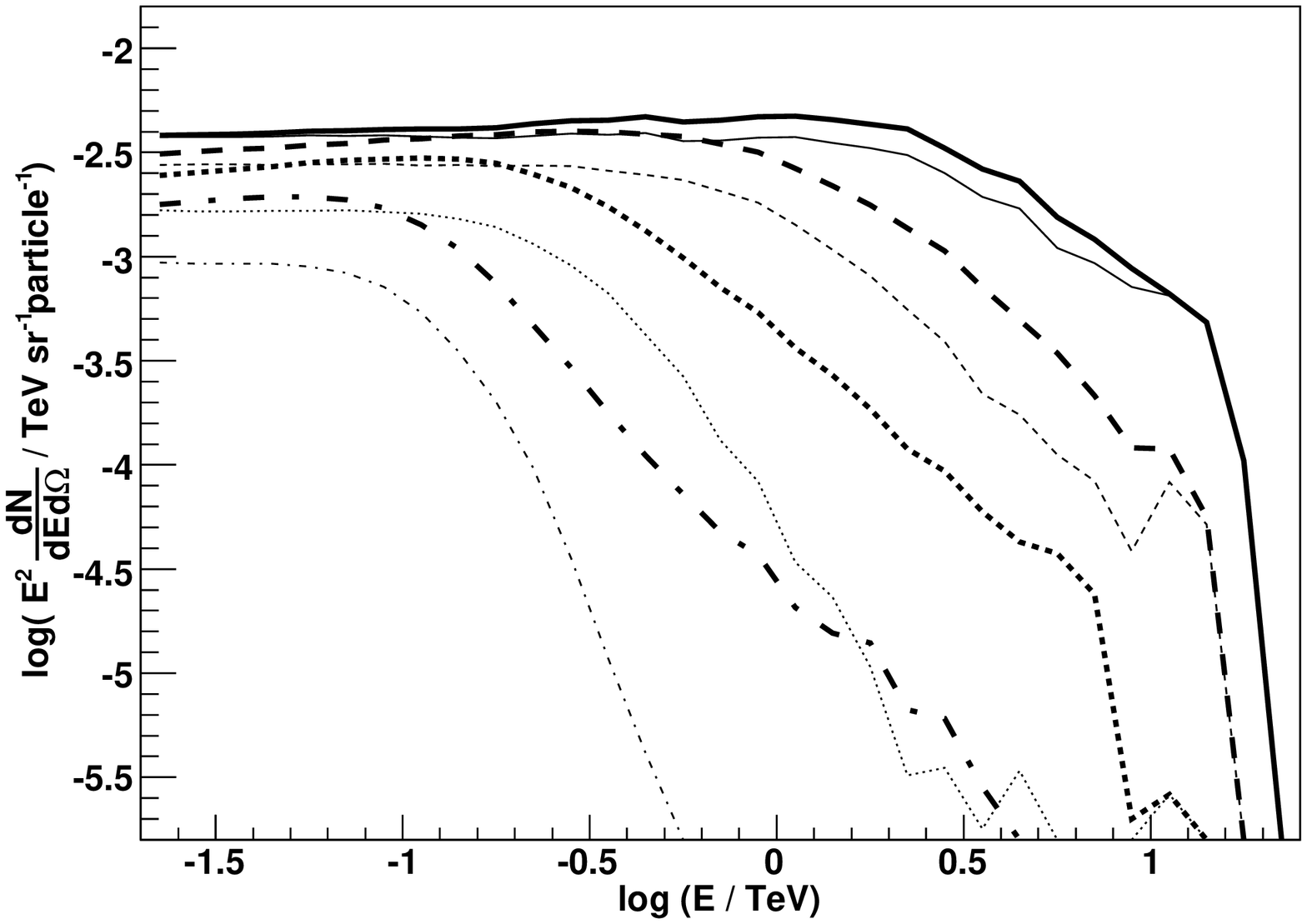}
\includegraphics[scale=0.28, trim= 32 26 0 0, clip]{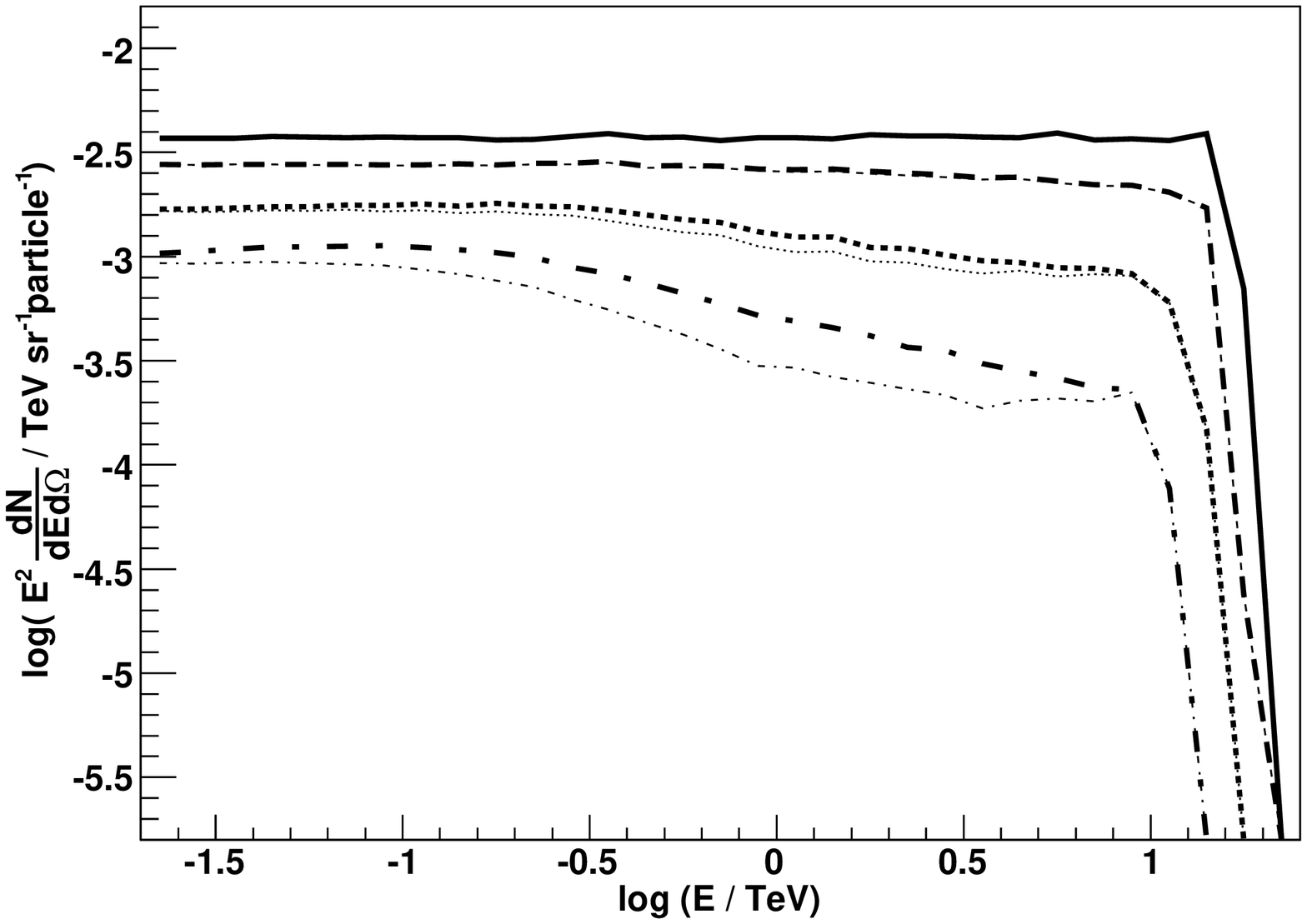}
\includegraphics[scale=0.28, trim= 0 0 0 0, clip]{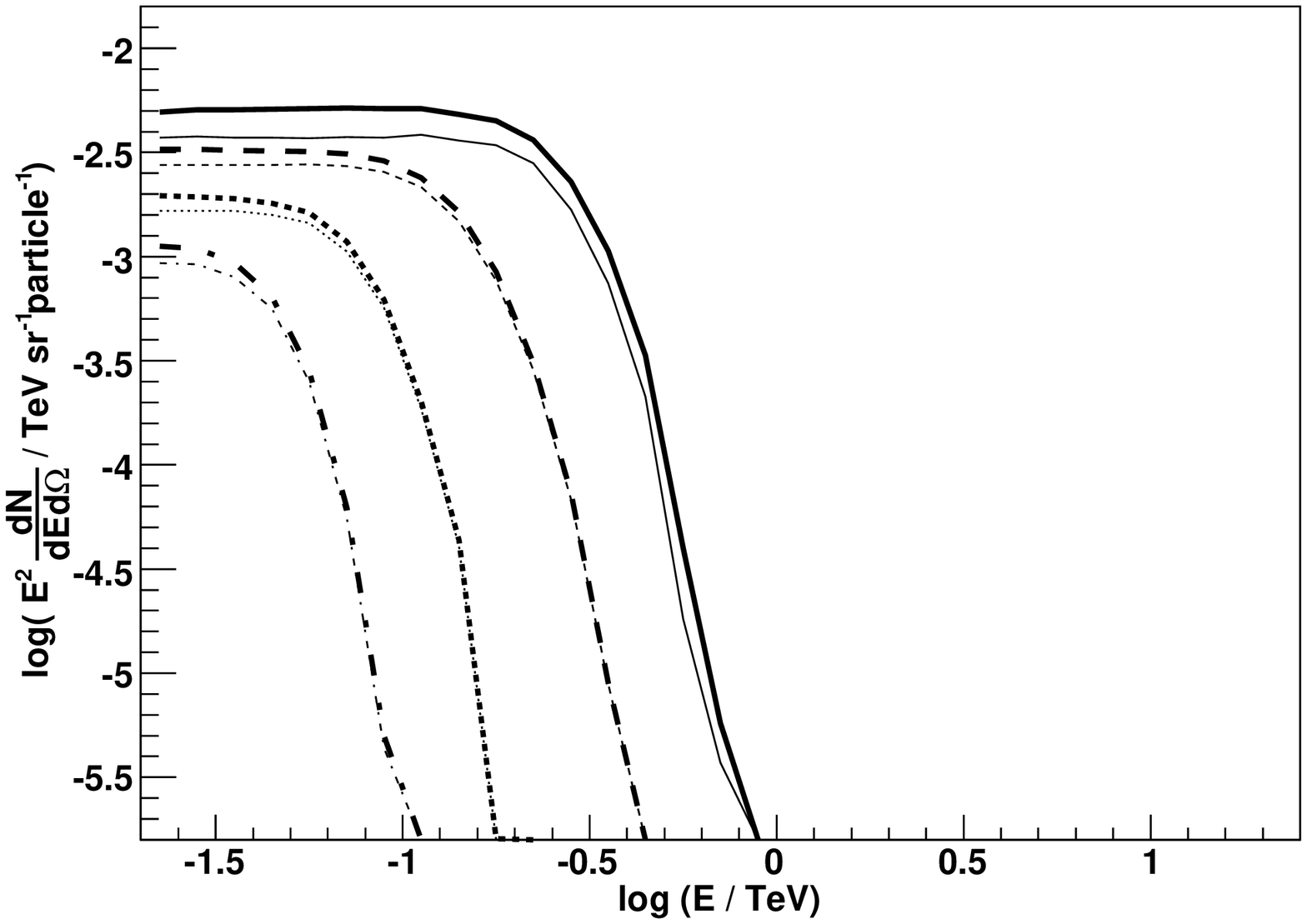}  
\includegraphics[scale=0.28, trim= 32 0 0 0, clip]{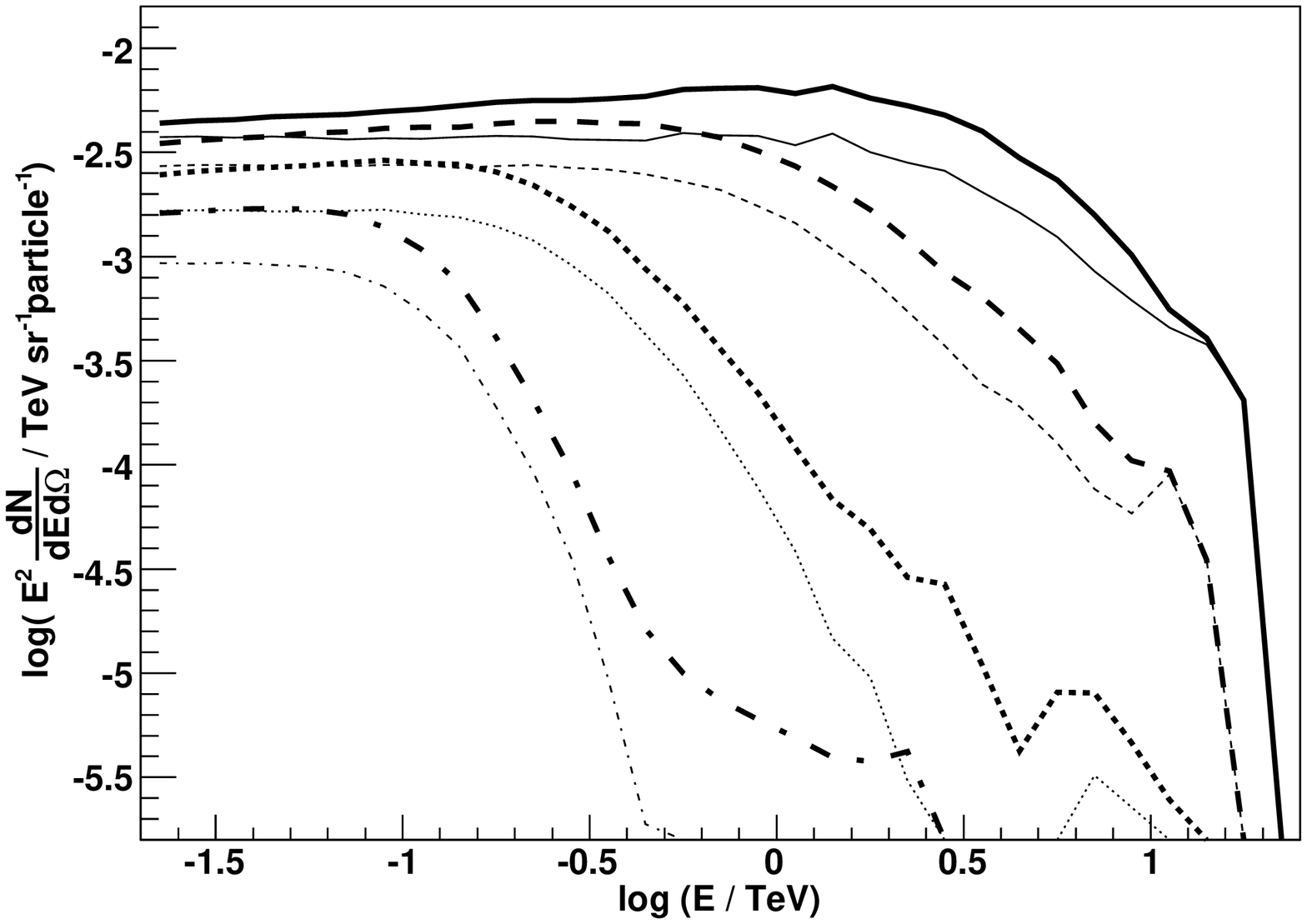}
\includegraphics[scale=0.28, trim= 32 0 0 0, clip]{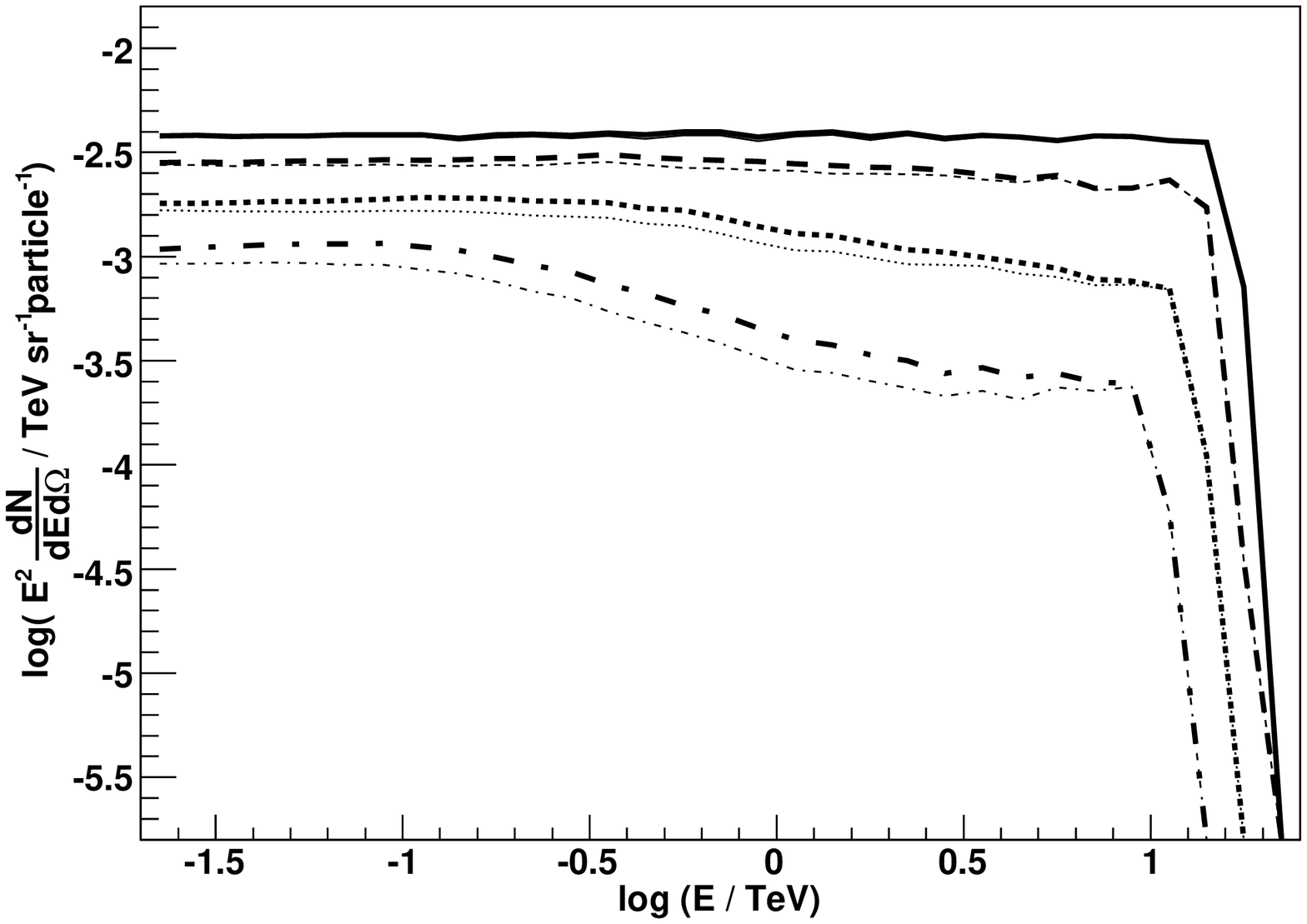}
\caption{Thick curves: $\gamma$-ray spectra produced in the IC $e^\pm$ pair cascade initiated by primary $\gamma$-rays injected isotropically from the point source (a blob) moving along the jet with the constant velocity $\beta = 0.5$ at a specific distances from the base of the jet: $H = 3r_{\rm in}$ (left figures) and  $30r_{\rm in}$ (middle), and $300r_{\rm in}$ (right). 
The spectrum of primary $\gamma$-rays is of the power law type with the differential spectral index $-2$.
The calculations have been performed without any magnetic field over the accretion disk (upper panel) and with the dipole magnetic field above the accretion disk which is normalized to $300$ G at the inner disk radius (bottom panel). 
Different line styles correspond to the escaping cascade spectra produced within the range of angle $\alpha$ measured in respect to the jet axis: 
$\alpha = 0^\circ - 20^\circ$ (solid), $20^\circ - 40^\circ$ (dashed),  $40^\circ - 60^\circ$ (dotted), and $60^\circ - 80^\circ$ (dot-dashed). 
Thin lines shows the primary, non-cascading component in the resulting spectrum. 
The resulting spectra are normalized to a single primary $\gamma$-ray per one steradian.}
\label{fig_spectra_fixed}
\end{figure*}

As an example we apply that the blob moves along the jet with velocity equal to $\beta = 0.5c$.
This results in a mild beaming of the emission in the direction of the jet.
The primary $\gamma$-rays have been injected from the blob with the power law spectrum above 10 GeV up to 10 TeV with the differential spectral index equal to $-2$. Their spectra are normalized to a single injected primary $\gamma$-ray within one steradian of the emission angle.
We transform energies and directions of the primary $\gamma$-rays to the disk reference frame and track the IC $e^\pm$ cascade initiated by them.
The $\gamma$-ray spectra emerging from the cascades initiated by primary $\gamma$-rays at a specific distance $H$ from the base of the jet at different range of observation angles $\alpha$ (measured in respect to the disk axis) are shown by the thick curves in Fig.~\ref{fig_spectra_fixed}. 
Note the dependence of the break in the cascade $\gamma$-ray spectra as a function of the distance of the blob from the black hole and the observation angle. In general, the break in the $\gamma$-ray spectrum 
shifts to higher energies for larger distance of the blob from the base of the jet.
For the blob at fixed distance from the disk, the cascade $\gamma$-ray spectra are steeper for the larger observation angles.

We also show the part of the escaping $\gamma$-ray spectra which are composed of the primary gamma-rays (see thin curves in Fig.~2). They were able to avoid absorption in the disk radiation. Let us discuss the main features of the $\gamma$-ray spectra 
escaping at specific angles.
In the case of an injection of the primary $\gamma$-rays far away from the accretion disk ($H \sim 300r_{\rm in}$), the absorption and also the cascading effects, are not very important. 
For small and moderate observational angles the primary spectrum is just modified by the term $e^{-\tau}$, with the cascading effects being negligible. 
However, even for those large injection distances, the cascading becomes evident at high observation angles, where the optical depths for $e^\pm$ pair absorption and IC scattering are still large enough for production of the next generations of $\gamma$-rays.

If the primary $\gamma$-rays are produced close to the source ($H \sim 3r_{\rm in}$), the whole cascade develops in the strong radiation field with many generations of particles being produced and absorbed.
This results in a strong cut-off in the escaping $\gamma$-ray spectrum.
The high energy photons are reprocessed in the cascade into lower energy photons, for which $\tau \ll 1$.
Therefore, the escaping cascade $\gamma$-ray spectra have a significant excess in the part of the spectrum below $50-100\mathrm{GeV}$, and also a slightly higher cut-off energy in respect to the primary $\gamma$-ray spectrum escaping to the observer. 

The $\gamma$-ray spectra in the intermediate distances from the black hole ($H \sim 30r_{\rm in}$) have interesting properties. 
Due to propagation of the primary $\gamma$-rays and $e^\pm$ pairs, the second generation of $\gamma$-rays is already created at the larger distances from the accretion disk, i.e. at places where the optical depths are lower.
So then, the second generation of particles will suffer lower absorption than the primary $\gamma$-rays. 
Therefore, the escaping $\gamma$-ray spectrum is significantly flatter at higher energies than the spectrum obtained from only primary $\gamma$-rays. 

We investigate also the role of the magnetic field above the accretion disk on the cascade $\gamma$-ray spectra.
The energy density of the dipole magnetic field falls more rapidly with a distance from a black hole then a energy density of the radiation field. 
In the calculations shown in the bottom panel of Fig.~\ref{fig_spectra_fixed}, we show the cascade $\gamma$-ray spectra for the case of the magnetic field which is in the equipartition with the energy density of the disk radiation (i.e. $B_{\rm in}~=~300$~G for $T_{\rm in} = 3\times 10^4$ K). 
The effect of the presence of the magnetic field above the disk is mostly important in the case of an injection relatively close to the base of the jet.
If the injection occurs at the distance $H \sim 3r_{\rm in}$, most of the energy of the $e^\pm$ pairs is radiated in the form of synchrotron radiation before IC interaction with a soft photon can occur.
This results in the full cascade spectra similar to the spectrum obtained from only primary, non-cascading $\gamma$-rays. 

\begin{figure*}
\centering
\includegraphics[scale=0.4, trim= 0  26 0 0, clip]{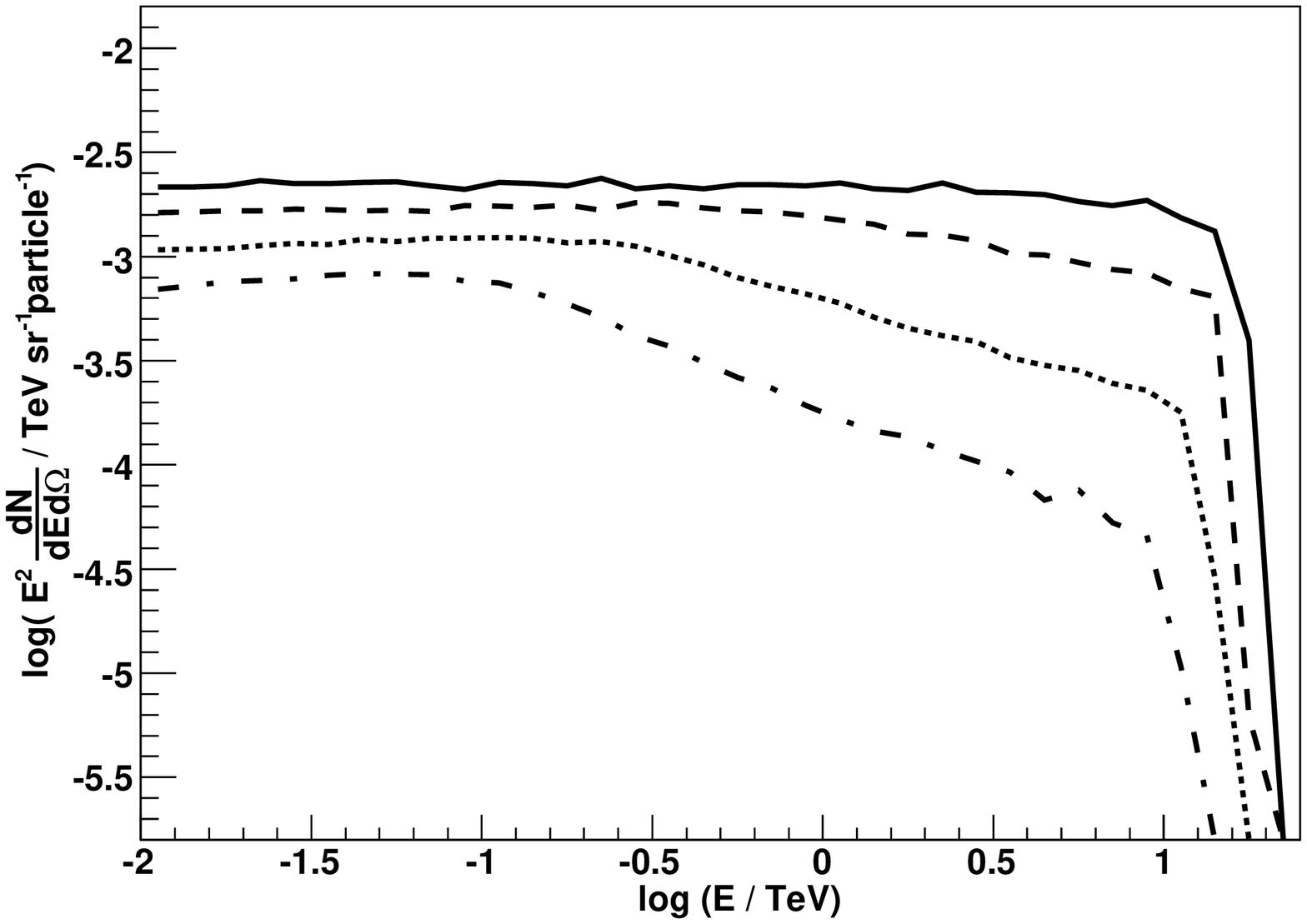}
\includegraphics[scale=0.4, trim= 32 26 0 0, clip]{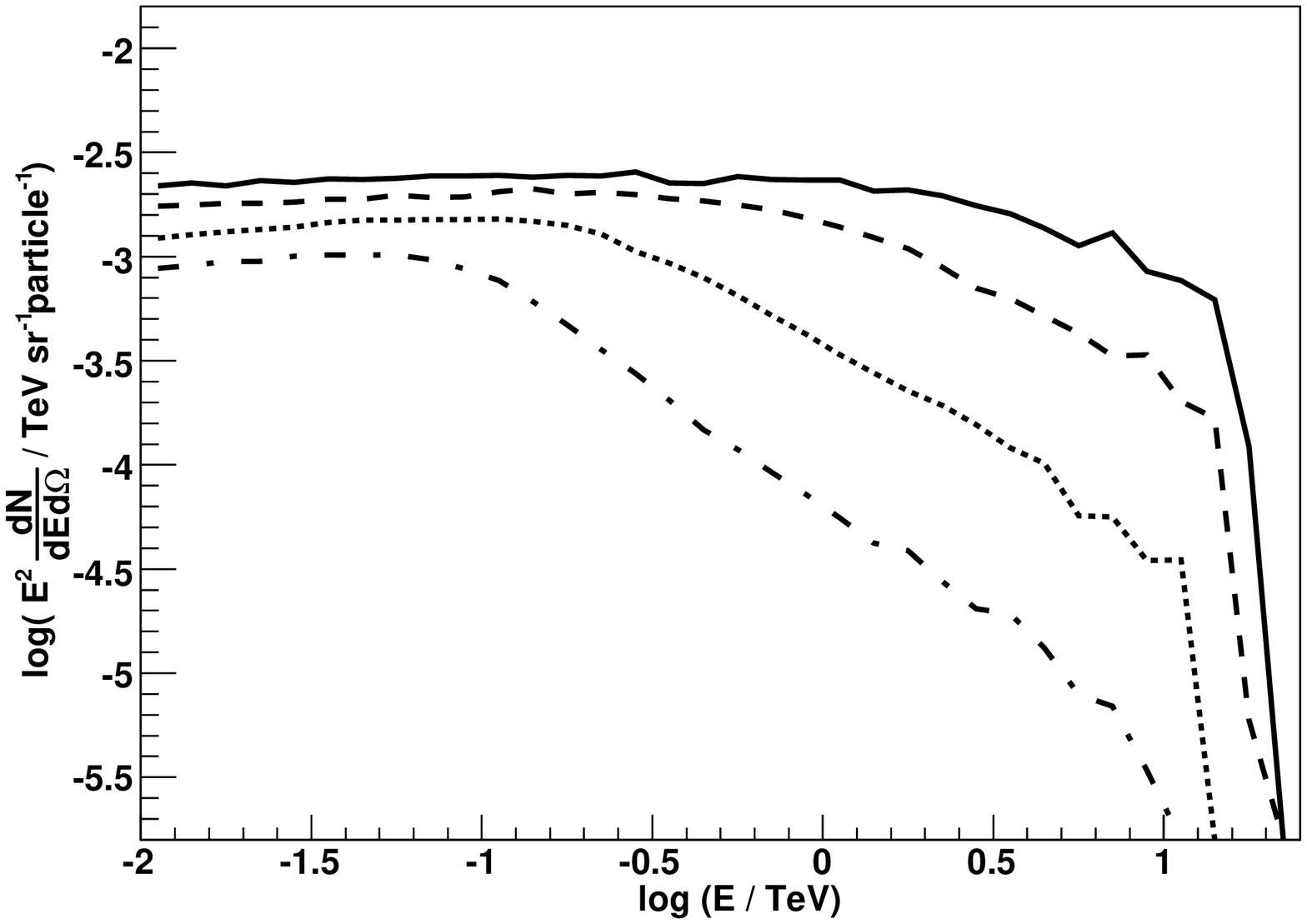}
\includegraphics[scale=0.4, trim= 0 0 0 0, clip]  {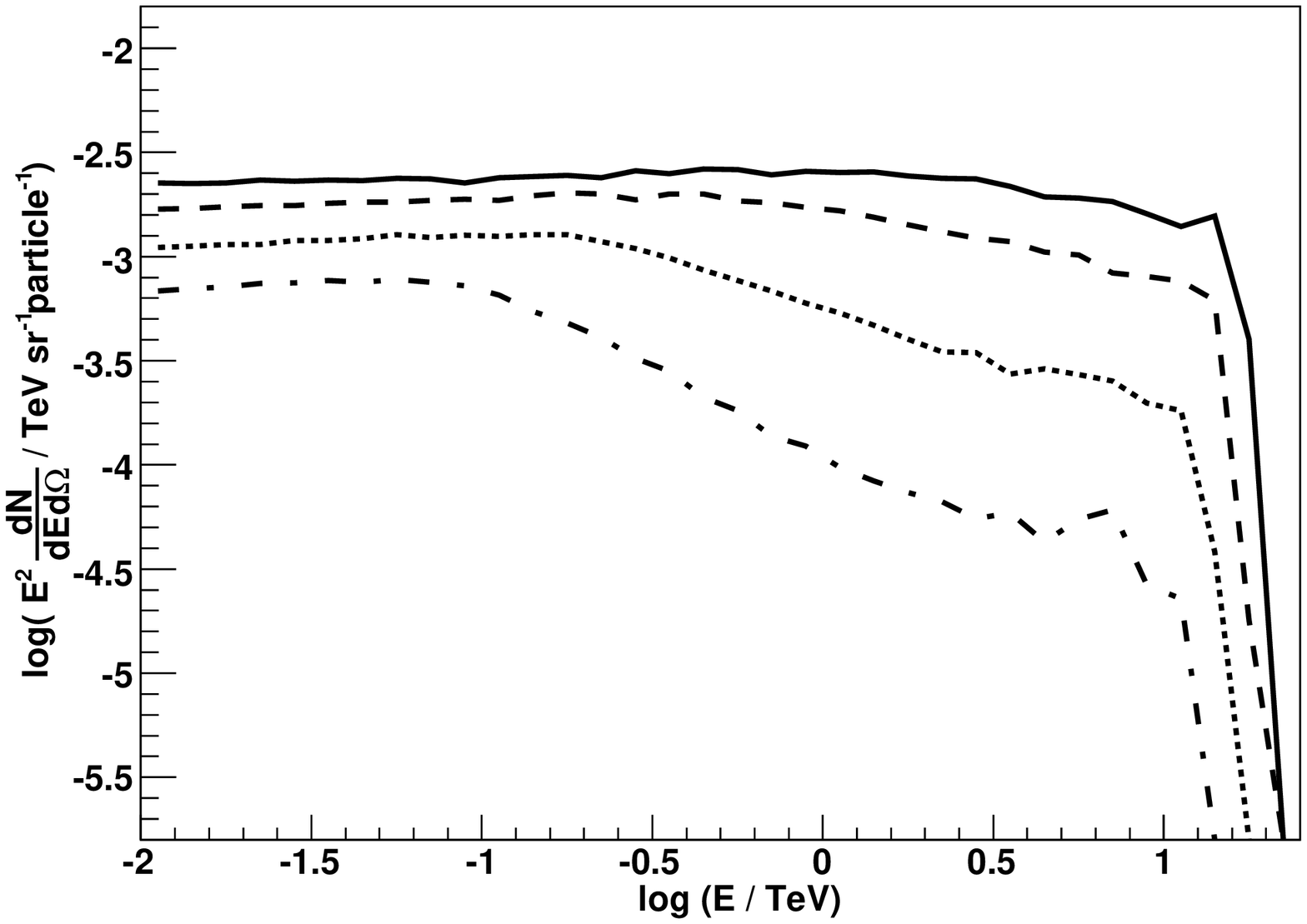}
\includegraphics[scale=0.4, trim= 32 0 0 0, clip] {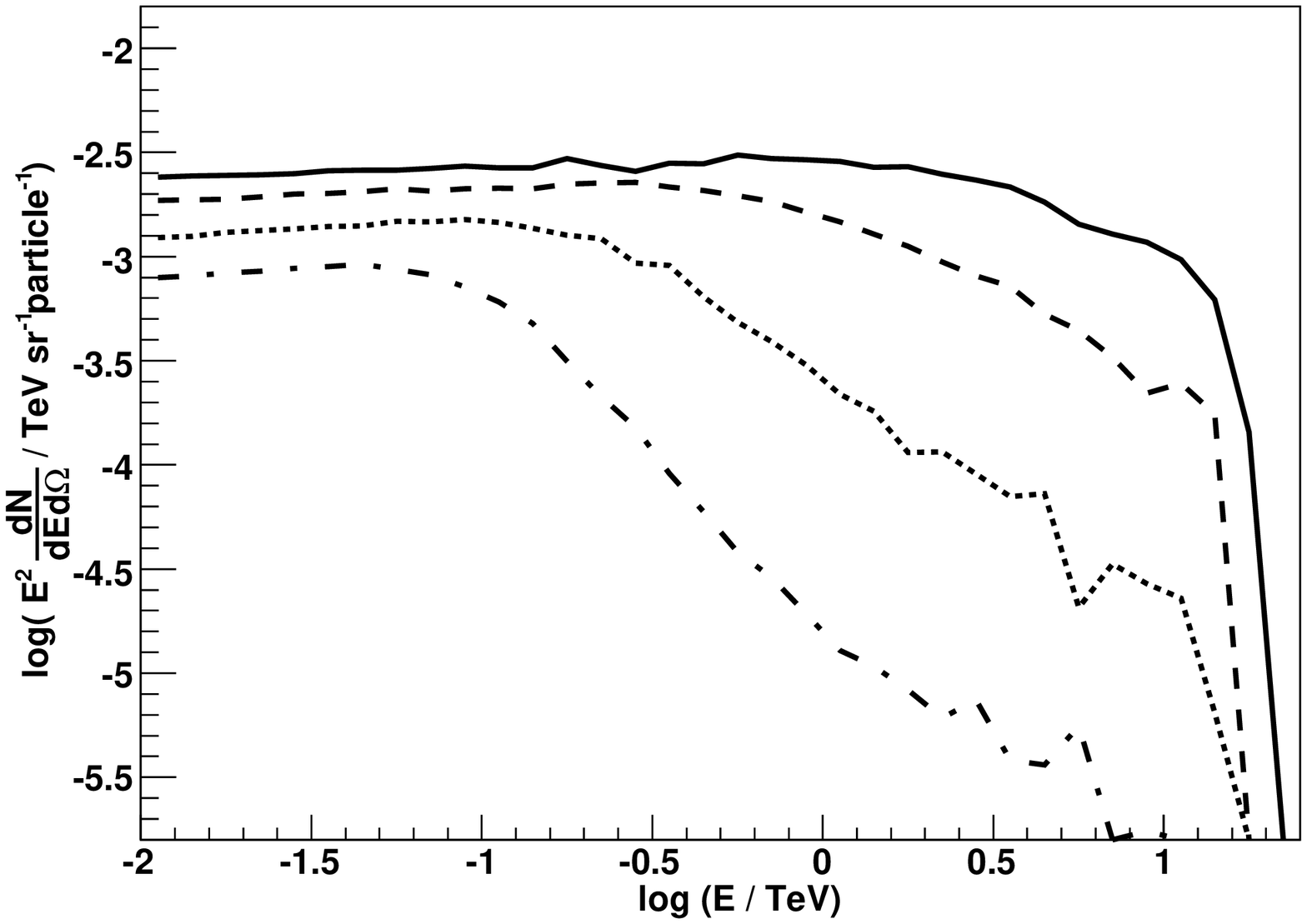}
\caption{$\gamma$-ray spectra produced in the $e^\pm$ pair cascade initiated by primary $\gamma$-rays injected isotropically from the point source moving along the jet with velocity $\beta = 0.5$ in the range of distances $H = 1 - 300 r_{\rm in}$ (left figures) and  $H = 1 - 100 r_{\rm in}$ (right). The spectrum of primary $\gamma$-rays is of the power law type with the differential spectral index -2. The calculations have been performed without any magnetic field above the accretion disk (upper panel) and with the dipole 
magnetic field normalized to $300$ G at the inner disk radius (bottom panel). The resulting spectra are normalized to a single primary $\gamma$-ray.
Line styles as in fig.~\ref{fig_spectra_fixed}.
}\label{fig_spectra_cont}
\end{figure*}
We also show the cascade $\gamma$-ray spectra produced by the blob injecting primary $\gamma$-rays as in the previous case but after integrating over the range of propagation distances
along the jet (see Fig.~\ref{fig_spectra_cont}). 
It is assumed that the blob moves through this part of the jet with fixed velocity $\beta = 0.5$ and injects primary $\gamma$-rays at a constant rate along the jet ($dN/dH\propto$ const).
As an example, we consider the cases of the blob moving from the base of the jet (at $H = r_{\rm in}$) up to the distance of $H = 100r_{\rm in}$ or  $300r_{\rm in}$. 
Note that the lower energy part of escaping $\gamma$-rays spectrum below a break is very similar to the injected spectrum of primary $\gamma$-rays (effects of the absorption and the cascading negligible).
However, for larger angles, the $\gamma$-rays spectra becomes steeper at the higher energy part. This part of the spectrum can be also well described by a simple power law  with the spectral index increasing with the observation angle. The break in the $\gamma$-ray spectrum (due to the cascade process) shifts to lower energies with the larger observation angle $\alpha$.  

\section{Discussion and Conclusion}

\begin{figure}
\centering
\includegraphics[scale=0.4, trim= 0  26 0 0, clip]{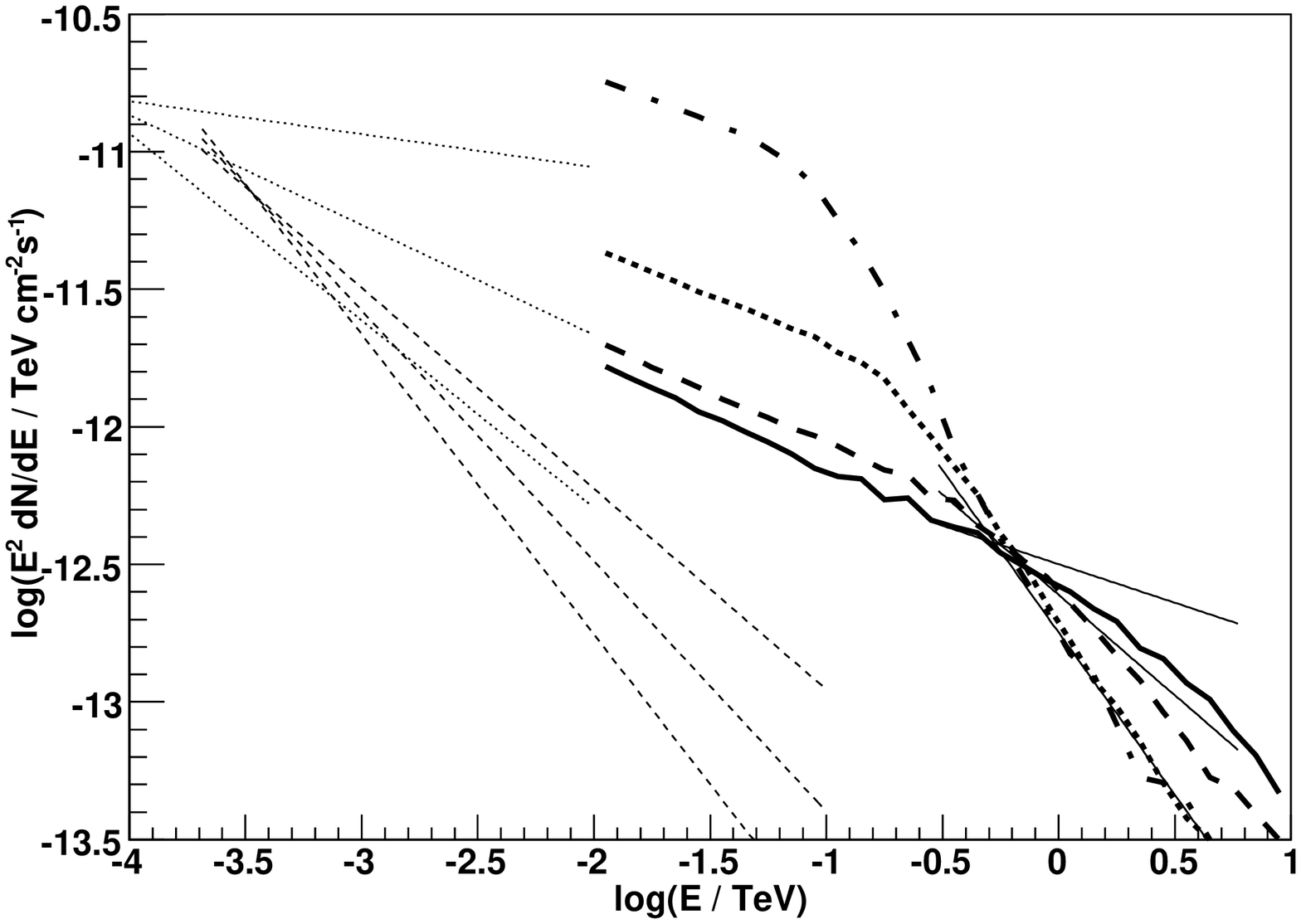} \\

\includegraphics[scale=0.4, trim= 0  0 0 0, clip]{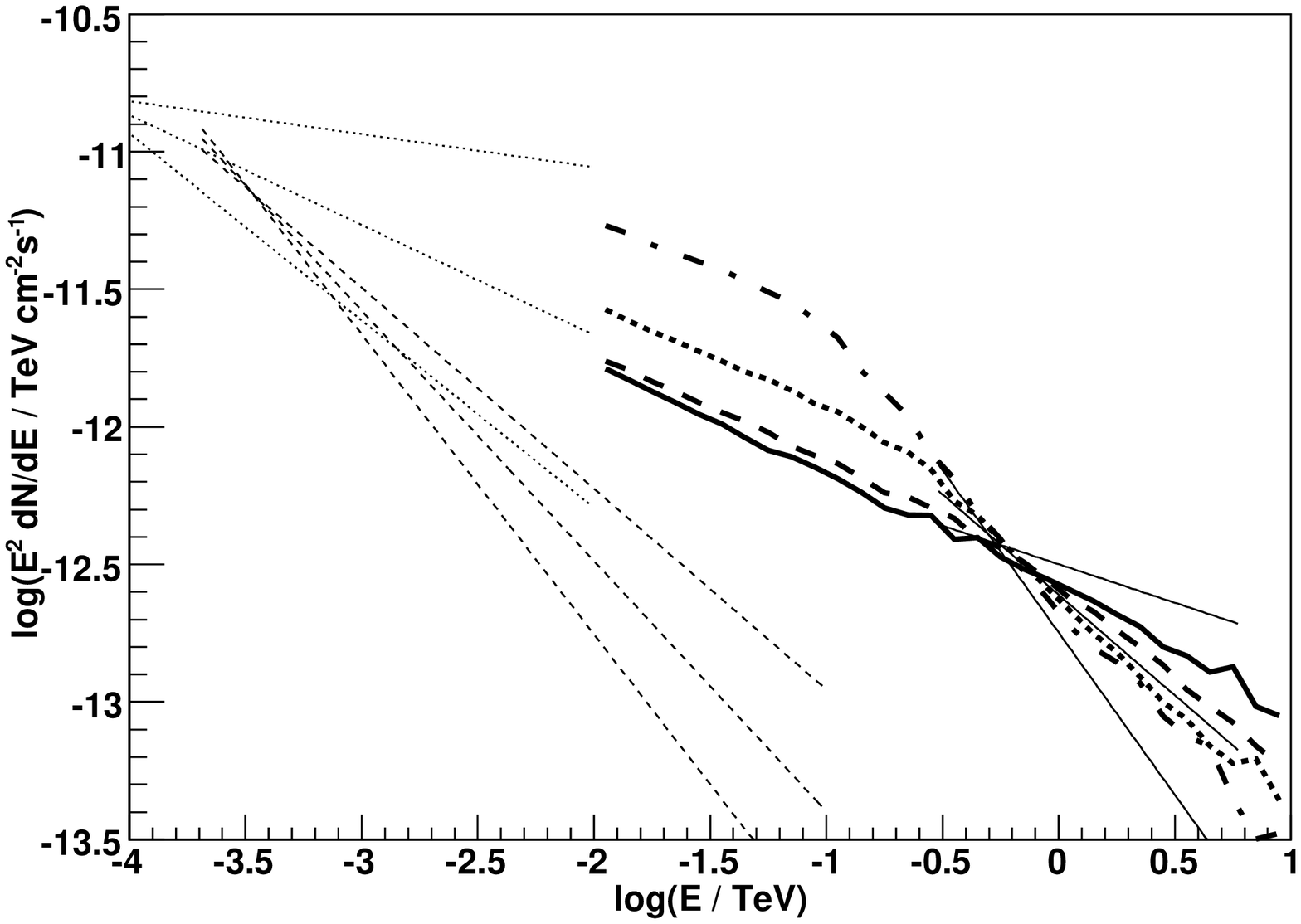}
\caption{
$\gamma$-ray spectra (together with an error estimation of the spectral index) measured by the EGRET (Sreekumar et al.~1999) (dotted), 
${\it Fermi}$ (Abdo et al.~2009) (dashed),
and H.E.S.S. telescopes (Aharonian et al.~2009) (solid) from Cen~A. 
They are compared with the $\gamma$-ray spectra produced in the IC $e^\pm$ pair cascade initiated by primary $\gamma$-rays injected isotropically from the blob moving along the jet with velocity $\beta = 0.5$ in the range of distances  $H = 1 - 100 r_{\rm in}$ (top figure) or $1 - 300 r_{\rm in}$ (bottom figure). 
The spectrum of primary $\gamma$-rays is of the power law type with the differential spectral index -2.4 (as observed in the GeV energies by EGRET). 
The calculations have been performed without any magnetic field above the accretion disk. 
The spectra are reported within the range of angle $\alpha$ measured with respect to the jet axis: 
$0^\circ - 20^\circ$ (thick solid curves), 
$20^\circ - 40^\circ$ (thick dashed),  
$40^\circ - 60^\circ$ (thick dotted), 
and $60^\circ - 80^\circ$ (thick dot-dashed).
Modeled spectra are normalized to have the same power as the H.E.S.S measurements integrated in the energy range of 300 GeV - 3 TeV.}
\label{fig_spectra_comp}
\end{figure}

We have calculated the $\gamma$-ray spectra expected in the case of an anisotropic
IC $e^\pm$ pair cascade developing in the disk radiation and magnetic fields in the whole volume above the accretion disk. 
The cascade is initiated by primary $\gamma$-rays produced in the blob moving along the jet.
These primary $\gamma$-rays might be produced in the mechanism internal to the blob (e.g. SSC or hadronic models). 
Our model is specially interesting in the case of close active galaxies observed at relatively large angles to the disk axis (often identified with the direction of the jet) at mildly relativistic speeds,
such as recently detected  GeV-TeV $\gamma$-ray radio galaxy Cen~A. 
We have performed cascade calculations for the supposed parameters of this source. 
As an example, we compare the $\gamma$-ray spectrum above $100$ MeV observed from Cen~A by the EGRET, {\it Fermi}, and H.E.S.S. observatories with the $\gamma$-ray spectra obtained in terms of our model (Fig.~\ref{fig_spectra_comp}).  The differential spectrum of primary $\gamma$-rays, injected from the blob into the disk radiation, has the spectral index equal to $-2.4$ as observed by the EGRET telescope
at the GeV energies. The calculations are done for two regions of $\gamma$-ray injection along the jet:
$H = 1-100r_{\rm in}$ and $H = 1-300r_{\rm in}$. The best fit to the observed spectrum is obtained for the inclination angle to the line of sight around $\alpha = 40^o$. This estimate is outside the range of values 
of the inclination of the jet ($\sim 50^o-80^o$) obtained on the basis of observed jet/counterjet brightness ratio (Tingay et al.~1998), but in the middle of the more recent limits  ($\sim 15^o-80^o$), derived by Hardcastle et al.~(2003) and Horiuchi et al.~(2006). Note that the comparison of the observed spectrum in the GeV-TeV energy range with results of calculations allows to constrain the extend of the emission region along the jet. For larger inclination angles production of $\gamma$-rays should occur farther from the black hole.
We conclude that the observed TeV $\gamma$-ray spectrum can be naturally explained in terms of this cascade model occuring in the whole volume above the accretion disk
for the injection region of primary radiation within the inner $\sim 100r_{\rm in}$
of the jet.

Note the interesting general features of the considered model. The $\gamma$-ray spectra escaping at different range of inclination angles have characteristic breaks in respect to the primary spectrum injected from the blob. The location of the break shifts to lower energies for larger inclination angles.
This feature might serve as diagnostic of the inclination of the accretion disk to to line of sight.  
   
Our model can be also applied to the blazars seen at lower inclination angles.
We expect that a part of the $\gamma$-ray emission produced in the cascade can contribute to the direct emission seen from the blob which escapes with negligable absorption along the jet axis. This $\gamma$-ray emission should have steeper spectrum. Moreover, it should be delayed in respect to the direct emission. As a result, the flare $\gamma$-ray emission observed from the specific blazar can exibit two components. Investigation of the flare component and the delayed component can provide information on the $\gamma$-ray production site in respect to the accretion disk. The interesting effects related to the prompt and delayed $\gamma$-ray emission will be studied in the future paper.

\section*{Acknowledgments}
This work is supported by the Polish MNiSzW grant N N203 390834.


\label{lastpage}

\begin{thebibliography}{99}
\bibitem{ab09} Abdo, A.A. et al. 2009 ApJ, 700, 597
\bibitem{ah07} Aharonian, F. et al. 2007 ApJ, 664, 71
\bibitem{ah09} Aharonian, F. et al. 2009 ApJ, 695, L40
\bibitem{al07} Albert, J. et al. 2007 ApJ 669, 862
\bibitem{be93} Bednarek, W. 1993 A\&A 278, 307
\bibitem{be97} Bednarek, W. 1997, MNRAS, 285, 69
\bibitem{be09} Bednarek, W. 2009, A\&A, 495, 919
\bibitem{bkm96} Bednarek, W., Kirk,J.G. 1995, A\&A, 294, 366
\bibitem{bp99} Bednarek, W., Protheroe, R.J. 1999, MNRAS 302,373
\bibitem{bl95} Blandford, R.D., Levinson, A. 1995, ApJ 441, 79
\bibitem{ca08} Cappellari, M. et al. 2009 MNRAS 394, 660
\bibitem{ca92} Carraminana, A. 1992 A\&A 264, 127
\bibitem{co92} Coppi, P.S. 1992, MNRAS, 258, 657
\bibitem{gh05} Ghisellini, G., Tavecchio, F., Chiaberge, M. 2005 A\&A 432, 401 
\bibitem{ha03} Hardcastle, M.J., Worrall, D.M., Kraft, R.P., Forman, W.R., Jones, C., Murray, S.S. 2003 ApJ 593, 169
\bibitem{ho06} Horiuchi, S. et al. 2006 PASJ 58, 211
\bibitem{is98} Israel, F.P. 1998 AAR 8, 237
\bibitem{le08} Lenain, J.-P., Boisson, C., Sol, H., Katarzy\'nski, K. 2008 A\&A, 478, 111
\bibitem{mb92} Mannheim, K., Biermann, P.L. 1992 A\&A 253, L21
\bibitem{mgc92} Maraschi, L., Ghisellini, G., Celloti, A. 1992 ApJ 397, L5
\bibitem{mhp95} Marcowith, A., Henri, G., Pelletier, G. 1995, MNRAS 277, 681
\bibitem{ne07} Neronov, A., Aharonian, F.A. 2007 ApJ 671, 85
\bibitem{pss83} Pozdnyakov, L. A., Sobol, I. M., Syunyaev, R. A. 1983, ASPRv, 2, 189
\bibitem{ri08} Rieger, F.M., Aharonian, F.A. 2008 A\&A 479, L5
\bibitem{ss73} Shakura, N.I., Sunyaev, R. 1973 A\&A 24, 337
\bibitem{sb08} Sitarek, J., Bednarek, W. 2008 MNRAS 391, 624
\bibitem{sr99} Sreekumar, P. et al. 1999 APh 11, 221
\bibitem{sv84} Svensson, R. 1984 MNRAS 209, 175
\bibitem{sv87} Svensson, R. 1987 MNRAS 227, 403
\bibitem{ti98} Tingay, S.J. et al. 1998 AJ 115, 960
\end{thebibliography}
\end{document}